\documentclass[9pt,conference]{IEEEtran}
\usepackage{waspaa25} 
\usepackage{bm} 
\usepackage{multirow,balance}
\usepackage{graphicx} 

\def\email#1{\gdef\@email{#1}}

\newcommand{\mymuspace}[1]{\mspace{#1mu}}

\title{Distributed Asynchronous Device Speech Enhancement\\ via Windowed Cross-Attention}
\name{
    $\mymuspace{8}$Gene-Ping Yang $\mymuspace{7}$
    Sebastian Braun $\mymuspace{8}$
}

\address{
    CSTR, University of Edinburgh $\mymuspace{7}$ Microsoft Research, Redmond $\mymuspace{8}$ \\
    $\mymuspace{55}$ geneping.yang@ed.ac.uk $\mymuspace{7}$ sebastian.braun@microsoft.com $\mymuspace{8}$
}

\begin{document}

\maketitle

\begin{abstract}

The increasing number of microphone-equipped personal devices offers great flexibility and potential using them as ad-hoc microphone arrays in dynamic meeting environments.
However, most existing approaches are designed for time-synchronized microphone setups, a condition that may not hold in real-world meeting scenarios, where time latency and clock drift vary across devices.
Under such conditions, we found transform-average-concatenate (TAC), a popular module for neural multi-microphone processing, insufficient in handling time-asynchronous microphones. 
In response, we propose a windowed cross-attention module capable of dynamically aligning features between all microphones. 
This module is invariant to both the permutation and the number of microphones and can be easily integrated into existing models. Furthermore, we propose an optimal training target for multi-talker environments.
We evaluated our approach in a multi-microphone noisy reverberant setup with unknown time latency and clock drift of each microphone.
Experimental results show that our method outperforms TAC on both iFaSNet and CRUSE models, offering faster convergence and improved learning, demonstrating the efficacy of the windowed cross-attention module for asynchronous microphone setups.

\end{abstract}

\section{Introduction}

While multi-device ecosystems are rapidly advancing and the number of personal devices brought to meetings is increasing, using multiple simultaneously active, asynchronous devices in the same acoustic environment is still a major challenge and poses awkard user problems.
Robust operation is often only guaranteed if the microphones and loudspeakers of all but one device in the same room are muted manually.
Ideally, robust advanced speech enhancement algorithms should be able to operate and deliver good audio quality without the user having to manually mute and unmute microphones and loudspeakers.
In this work, we focus on the voice pickup side, i.e. the microphone signal processing side of this problem. 
Main challenges causing existing speech enhancement methods to fail are that 1) the microphone and speaker positions can change dynamically, 2) each device is running on its own clock, which can cause clock drift between audio streams digitized on different devices \cite{gaubitch2013auto,wang2015self,cherkassky2017blind,bahari2017blind,hu2022distributed,schmalenstroeer2017multi}, and 3) the transmission over typical network protocols (i.e. the internet) can introduce large latency between streams of over 100~ms. 

All existing speech enhancement techniques designed for such "ad-hoc microphone array" scenarios ignore at least one of these conditions or depend on error-prone online synchronization techniques and therefore fail in practice \cite{9003849,9054177,luo21c_interspeech,wang2018multichannel,kim24m_interspeech,zhang2021,9746876,pandey22c_interspeech,6694323,tavakoli2016framework}.
Many methods assume perfectly synchronized microphone signals, which is essential for traditional beamforming \cite{capon1969high,griffiths1982alternative, souden2009optimal, gannot2017consolidated}.
However, ad-hoc arrays, which use microphones spread across different devices and operating without synchronization, typically lack this coordination, leading to poorer performance \cite{zhang2021}.
In such setups, simple microphone mixing can introduce issues like comb-filtering or echo effects.
Moreover, approaches like \emph{best microphone selection} are suboptimal in multi-speaker environments, especially when several speakers are active simultaneously and are prone to fast microphone switching artifacts \cite{araki2018,kumatani2011channel,wolf2014channel,zhang2020study,yoshioka2022picknet,wang2021continuous,8234698,zhang2021sensor}.
Clock drift between different recording devices is often overlooked in many methods, which can cause gradual temporal misalignment of the signals over time \cite{miyabe2015blind,gaubitch2013auto,wang2015self}.
While these issues could be resolved using heuristics, such as cross-correlation based alignment \cite{yoshioka2019meeting} and potentially dynamic time warping to address clock drift, the error propagating from misaligned signals can degrade overall system performance, and reaction time to changing conditions is typically too slow.
PickNet \cite{yoshioka2022picknet} proposes a mic selection method, which however assumes a single active speaker and assumes the delay to be within the same frame. SAMbA \cite{furnon2023samba} propose an attention-based mic stream merging, but treat delay and clock drift separately and do not look into multi-talker scenarios.
In response to these challenges, we propose a novel end-to-end model that effectively synchronizes and enhances asynchronous microphone streams for multi-talk scenarios, eliminating the need for potentially erroneous cascaded modules.

Our proposed method builds upon previous efforts on neural beamforming \cite{9054177,yoshioka2022picknet,9746876,guo24_interspeech, ristea23_interspeech} by introducing an efficient temporal cross-device attention module, called Windowed Cross-Attention (WCA), which is specifically designed to adaptively perform temporal alignment in distributed asynchronous microphone setups. 
WCA leverages cross-attention to dynamically align and aggregate information across microphones, accounting for device latency and clock drifts by operating within a fixed temporal window. 
This allows for more efficient information sharing between audio streams while minimizing the required memory complexity, making it suitable for real-time applications. 
Unlike previous methods, WCA is capable of processing asynchronous inputs, ensuring robust performance even in the presence of dynamically varying microphones and mobile speakers. 
We further investigate different training targets to deal with multiple and simultaneously active talkers.
We implement three strategies: random mic, minimum latency and closest microphone per speaker.

We evaluate the proposed WCA module on simulated datasets with background noise, multi-talk, stream delays and clock drift.
We validate the effectiveness of the WCA module by incorporating it into two backbones: implicit Filter-and-Sum Network (iFaSNet) \cite{luo21c_interspeech} and Convolutional Recurrent U-Net (CRUSE) \cite{9413580, 10096131}.
We explore different strategies for training target selection, such as random selection, closest microphone (CD), and minimum latency.
Our results indicate that WCA is a versatile module that generalizes well across architectures like iFaSNet and CRUSE, can be integrated into a range of multi-channel speech enhancement systems, and generalized well to real-world conditions.

\section{Problem Formulation}

A set of $M$ microphones captures a mixture of noisy speech signals from $S$ speakers in a reverberant environment. Let $x_m(t)$ denote the signal received at microphone $m$ and is modeled as:
\begin{equation}
    x_m(t) = \sum_{k=1}^S h_{m,k}(t) * s_k(t) + n_m(t), \quad m = 1, ..., M
\end{equation}
where $h_{m,k}(t)$ is the room impulse response (RIR) between speaker $k$ and microphone $m$, 
$s_k(t)$ is the clean speech signal of speaker $k$,
$n_m(t)$ is the additive noise at microphone $m$, 
and $*$ denotes the convolution operation.
The actual observation vector $X$ may be asynchronous, i.\,e.\, the microhpone signals are not time-synchronized. 
As each microphone is on a separate device, each can have an individual microphone latency and clock drift. 
The microphone observation $X$ is then given by 
\begin{equation}
    X \approx \left[ x_1(\gamma_1 t -\tau_1), ..., x_m(\gamma_m t -\tau_m), ... \right],
\end{equation} 
where $\tau_m$ denotes the latency of microphone $m$, which result from processing delays or network-induced latency, and 
$\gamma_m t$ captures the clock drift, which accumulates over time.

The objective is to estimate the sum of clean speech signals, which are potentially delayed and scaled by a dirac-delta function $\delta_k$:
\begin{equation}
    y(t) = \sum_{k=1}^S \delta_k(t) * s_k(t).
\end{equation}
The function $\delta_k(t)$ is related to a direct-path impulse response, which depends on the chosen training target discussed in Sec.~\ref{sec:target_design}.

\begin{figure}[tb]
    \centering
    \includegraphics[width=0.75\columnwidth,clip,trim=280 200 300 150]{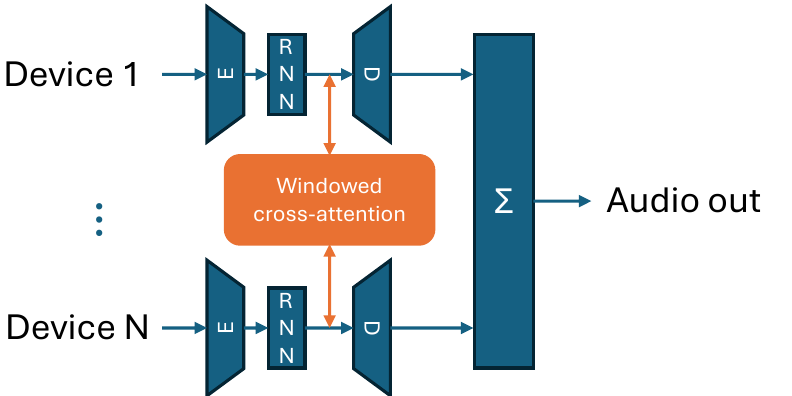}
    \caption{Proposed system with U-Net per stream and WCA connections after RNN bottleneck.}
    \label{fig:blockdiag}
    \vspace{-0.8em}
\end{figure}
\section{Related Work}

We review two recent studies that address the multi-channel microphone setups, where both are designed under the assumption of time-synchronized inputs \cite{9054177,guo24_interspeech}.
In transform-average-concatenate (TAC), a pooling layer is applied independently at each time step across the hidden features of all microphones for a beamformed representation \cite{9054177}. 
Given microphone representations $Z_m = f(X_m)$, where $f(\cdot)$ denotes an encoder network, TAC can be expressed as:
\begin{align}
    A_t = \frac{1}{m}\sum_m F_{m, t}, \quad 
    F_{m,t} = P(Z_{m,t})
\end{align}
where $P$ is a linear projection layer, and only frames with the same time index $t$ are averaged.
This same aggregated information $A_t \in \mathbb{R}^d$ is shared across all microphones and combined with each $Z_{m,t}$.

Another concurrent work is the self-attention aggregation module \cite{guo24_interspeech}.
The term self-attention may be misleading, as the attention operates along the microphone axis rather than the time axis.
The microphone aggregation is formulated as:
\begin{equation}
    A_{t} = softmax(\dfrac{Q_t \cdot K_t^\top}{\sqrt{d}}) \cdot V_t
\end{equation}
where $A_t, Q_t, K_t, V_t \in \mathbb{R}^{M \times d}$ and $d$ denotes the hidden dimension.
Similar to TAC, this approach restricts attention to frames with the same time index $t$ across microphones, limiting its ability to handle temporal misalignment across microphones.

These two approaches can be effective when the synchonization offset between microphones remains within a single processing frame. 
For instance, if input spectra are extracted using a 20 ms window, delays within this range may still align to the same time index $t$. 
However, they do not account for temporal misalignment larger than one analysis window, often leading to failure in practice.

\section{Proposed system}
We build our system on a single-channel speech enhancement network, consisting of encoder-bottleneck-decoder modules. We propose to add a cross attention module to communicate between the per microphone channel modules to allow synchronization and extract the optimal microphone. A system overview is shown in Fig.~\ref{fig:blockdiag}.

\subsection{Temporal Cross-Attention}

We propose to align microphone features using a temporal cross-attention module.
Our proposed module operates on the hidden representations of microphone signals similar to previous methods \cite{9054177,luo21c_interspeech,guo24_interspeech}.
From the hidden representations of all microphone signals $Z \in \mathbb{R}^{M \times T \times d}$, where $T$ is the number of encoded frames, Query, Key and Value are first derived through linear projection:
\begin{equation}
    Q_{m} = P_Q(Z_{m}),\; K_{m} = P_K(Z_{m}),\; V_{m} = P_V(Z_{m}), 
\end{equation}
where $P_Q, P_K,P_V$ are linear projection layers to dimension $d$.
Next, for each microphone $m$, a cross-attention layer is performed to aggregate information from every microphone $n$ to microphone $m$:
\begin{equation}
    A_m = \sum_n softmax(\dfrac{Q_m \cdot K_n^\top}{\sqrt{d}}) \cdot V_n
    \label{equ:cross_attention}
\end{equation}
where $A_m, Q_m, K_n, V_n$ are all with shape $\mathbb{R}^{T \times d}$.
The dot product $(Q_m \cdot K_n^\top)$ computes pairwise similarity scores between all time frames of microphone $m$ and microphone $n$.
A subsequent softmax then determines the relative importance of each time frame in $K_n$ with respect to a specific frame in $Q_m$.
This process can be interpreted as a neural beamforming technique, effectively aligning all microphone signals with the target microphone $m$.
Finally, we concatenate the beamformed representation with the original input $Z_m$ and apply a linear transformation to produce the final output:
\begin{equation}
    \hat{Z}_m = P_c(concat[Z_m, P_A(A_m)])
\end{equation}
where $P_A$ is a projection layer applied to the aggregated microphone-specific representation $A_m$, and $P_C$ is a linear layer that refines the combined features.

\subsection{Windowed Cross-Attention}

In the previous cross-attention formulation, every microphone pair $(m, n)$, requires an attention matrix of size $T \times T$, resulting in a memory requirement of $O(M^2T^2)$ in total.
This significantly increases memory requirements for longer signals and is unfeasible in a streaming fashion.
Referring back to Equation \ref{equ:cross_attention}, we define the similarity matrix as $S_{i,j} = Q_m \cdot K_n^\top$ where $i$ and $j$ represent the frame indices for microphone $m$ and $n$ respectively.
However, it is highly unlikely that frame $i=0$ from microphone $m$ would align with frame $j=T$ from microphone $n$, given that inter-microphone latency is typically constrained within a limited range ($\Delta \tau \ll T$).

Therefore, we propose restricting the alignment of frame $i$ to frame $j$ where $i-L \leq j \leq i+L$, and $L$ represents the expected maximum time offset between two microphones.
To achieve this, we first unfold $K_n$ and $V_n$ along the time axis, transforming them into $K_n^u, V_n^u \in \mathbb{R}^{T \times (2L+1) \times d}$, where each $K_n^u[i]$ consists of a local window $K_n[i-L:i+L+1]$, and the same for $V_n^u$.
We then reformulate the attention computation as follows:
\begin{equation}
    A_{m}[i] = \sum_n softmax(\dfrac{Q_m[i] \cdot K_n^u[i]^\top}{\sqrt{d}}) \cdot V_n^u[i]
\end{equation}
where $Q_m[i]$ attends only to a local window of $K_n$, effectively constraining the attention to a limited temporal range.

\subsection{Output Target Design}
\label{sec:target_design}

The training target in multi-microphone speech enhancement model is chosen as the clean speech at a fixed, user-defined microphone channel, typically the first channel \cite{luo21c_interspeech,9054177,9003849, wang2018multichannel, kim24m_interspeech}. 
However, without prior knowledge and for dynamic scenes, such a random microphone as the target may be suboptimal, as it does not necessarily correspond to the microphone closest to the active speaker. 
A more effective approach would be to use the closest microphone to the active speaker as the target, similar to the strategy proposed in PickNet \cite{yoshioka2022picknet}.
However, PickNet does not account for overlapping speech and only classifies the single best microphone at each time frame $t$.
In simultaneous multi-talk scenarios, this approach may be limited and suboptimal.

To address this limitation, we design three different target selection strategies for training the model.
\begin{enumerate}
    \item \underline{Random Microphone Selection}: As in most prior work, we use the direct sound mixture from at a priori fixed, randomly chosen microphone as the target to maintain comparability with existing approaches.
    \item \underline{Minimum Latency Selection}: We select the microphone with the minimum latency $m^* = \arg\min_m \tau_m$ ensuring minimal delay in capturing the direct clean speech signal.
    \item \underline{Closest Microphone to Each Speaker}: We select the microphone closest to each active speaker and mix the corresponding direct path speech signals to form the training target. This is formulated as: 
    \begin{equation}
        y(t) = \sum_{k=1}^S (h_{m_k, k}^{\text{d}} * s_k) (\gamma_{m_k}t - \tau_{m_k})
    \end{equation}
    where $m_k$ is the microphone closest to speaker $k$ and $h_{m_k, k}^{\text{d}} (t)$ is the direct-path of the RIR.
\end{enumerate}
\section{Experiments}

\subsection{Dataset}

We conduct our experiments on large-scale synthetic training, validation, and testing from real recordings, where different datasets are used for each stage.
For data simulation, we follow the pipeline outlined in \cite{9413580,10096131}, leveraging publicly available speech and noise datasets. 
The training data consists of clean speech from LibriVox and VoxCeleb2 and noise from the DNS challenge \cite{dnschallenge2023}, with 5,120 training samples of 10~s length generated on-the-fly per epoch. 
The validation set is constructed using clean speech from VCTK and DEMAND, resulting in a total of 300 synthetic mixtures. 
The test set consists of 1,000 mixtures, sampled from DAPS and QUT datasets.

For each simulated sample, the number of speakers is randomly selected between 1 and 3, while the number of microphones varies between 1 and 6.
The speaker overlap ratio is set to approximately 50\% to ensure a balanced mix of overlapping and non-overlapping speech segments.
To simulate reverberation, speech and noise signals are augmented with RIRs generated using the image-source method \cite{Allen1979}. Background noise is simulated as spatially diffuse by summing 64 point noise sources spread across the room.
The reverberant speech and noise is mixed with a signal-to-noise ratio (SNR) drawn from a Gaussian distribution with $\mathcal{N}(5,10)$ dB. 
The resulting mixture signals are re-scaled to levels distributed with $\mathcal{N}(-40,10)$ dB.

To simulate asynchronous device conditions, we introduce device latency by applying random time shifts to each microphone, with delays sampled between -40 ms and 40 ms. 
Additionally, clock drift is simulated using a Gaussian distribution with a mean sample rate of 16 kHz and a standard deviation of 0.5, ensuring realistic temporal misalignment across microphones.

\subsection{Model Configurations}

We evaluate our proposed Windowed Cross-Attention (WCA) by modifying two established speech enhancement backbones. 
First, we adapt the multi-microphone iFaSNet model \cite{luo21c_interspeech}, by substituting its standard TAC module with our WCA.
The model consists of 6 DPRNN blocks \cite{luo2020dual}, each with a hidden dimension of 128, and each block is followed by a WCA module.
Since the hidden representation of every microphone after each DPRNN block is of shape $\mathbb{R}^{C \times W \times d}$, with $C$ representing the number of chunks and $W$ the chunk length, we convert it back to the sequential format of shape $\mathbb{R}^{T \times d}$ using overlap-add before being forwarded to WCA.
After processing through WCA, we re-segment the hidden representation into its 3D format with overlapping chunks.
We set the window length to 20 ms and a 50\% overlap ratio to process the encoded input features.
Our second model adapts the single-channel speech enhancement model CRUSE \cite{Braun2022}, by incorporating a WCA module after the bottleneck layer to aggregate multi-channel information. 
The model setup follows \cite{Braun2022,9413580}, a U-Net architecture of 4 symmetric encoder and decoder layers, with encoder filters configuration of [32, 64, 64, 64]. 
The STFT window length is set to 20 ms with a 50\% overlap ratio, and a single GRU layer is placed at the bottleneck, followed by the WCA module. We modify CRUSE to directly predict the complex compressed spectrum instead of a filter to be able to produce time-shifts.
For both iFaSNet and CRUSE, we set $L$ to 4 in WCA, i.e., a cross-attention window of 90 ms.
The final output is the sum of the output signals across all channels.
Models are trained using the complex compressed MSE loss \cite{braun2021consolidated}, with a batch size of 64 and a learning rate of 0.001.

\begin{table}[t!]
    \caption{Evaluation on the full test set using DNSMOS (SIG, BAK, OVRL), XLSR-MOS and Cepstral Distance (CD).}
    \label{tab:full-test-set}
    
    \begin{center}
    \begin{tabular}{ccc|ccccc}
    {\bf} & {\bf\footnotesize Module} & {\bf\footnotesize Target} & {\bf\footnotesize SIG} & {\bf\footnotesize BAK} & {\bf\footnotesize OVRL} & {\bf\footnotesize XLSR} & {\bf\footnotesize CD} \\ 
    \toprule
    \multirow{6}{*}{\rotatebox{90}{iFaSNet}} 
     & TAC  & rand & 2.68 & 3.72 & 2.28 & 2.11 & 4.6 \\
     & WCA  & rand & 2.67 & {\bf 3.80} & 2.27 & {\bf 2.12} & 4.5 \\
    \cmidrule{2-8}
     & TAC  & min\_$\tau$ & 2.55 & 3.34 & 2.09 & 1.91 & 5.2 \\
     & WCA  & min\_$\tau$ & 2.55 & 3.76 & 2.14 & 1.89 & 4.7 \\
    \cmidrule{2-8}
     & TAC  & CM & 2.38 & 2.65 & 1.94 & 1.97 & 5.6 \\
     & WCA  & CM & {\bf 2.74} & 3.75 & {\bf 2.35} & 2.10 & 4.5 \\
    \midrule
    \multirow{6}{*}{\rotatebox{90}{CRUSE}} & TAC  & rand & 2.43 & 3.31 & 1.98 & 2.08 & 5.1 \\
     & WCA  & rand & 2.63 & 3.71 & 2.22 & 2.14 & 4.6 \\
    \cmidrule{2-8}
     & TAC  & min\_$\tau$ & 2.55 & 3.66 & 2.04 & 1.70 & 6.3 \\
     & WCA  & min\_$\tau$ & 2.63 & {\bf 3.79} & 2.24 & 2.24 & 4.7 \\
    \cmidrule{2-8}
     & TAC  & CM & 2.33 & 2.62 & 1.92 & 2.15 & 5.3 \\
     & WCA  & CM & {\bf 2.78} & 3.75 & {\bf 2.41} & {\bf 2.35} & 4.9 \\
    \midrule
    \multicolumn{3}{l|}{\footnotesize Rand. Mic (Noisy)} & 2.02 & 2.04 & 1.58 & 2.02 & - \\
    \multicolumn{3}{l|}{\footnotesize Single-ch + Rand. Mic} & 2.71 & 3.81 & 2.34 & 2.23 & - \\
    \multicolumn{3}{l|}{\footnotesize Single-ch + Picknet} & 2.72 & {\bf 3.83} & 2.35 & 2.11 & - \\    
    \bottomrule
    \end{tabular}
    \end{center}
    \vspace{-1.2em}
\end{table}

\begin{table*}[ht!]
    \caption{Ablation study comparing the performance of different modules (TAC vs. WCA) and various input conditions (varied $\gamma$, $\tau$, and speaker overlap) using the CRUSE model. Performance is reported using OVRL from DNSMOS and XLSR-based MOS.}
    \label{tab:ablation}
    
    \begin{center}
    \begin{tabular}{ccccccccccccc}

    & & & \multicolumn{2}{c}{\footnotesize $\gamma = 1$} & \multicolumn{2}{c}{\footnotesize std($\gamma$) = 0.5} & \multicolumn{2}{c}{\footnotesize std($\gamma$) = 2} & \multicolumn{2}{c}{\footnotesize Non} & \multicolumn{2}{c}{\footnotesize Full} \\
    
    & & & \multicolumn{2}{c}{\footnotesize $\Delta \tau = 40$ms} & \multicolumn{2}{c}{\footnotesize $\tau = 0$} & \multicolumn{2}{c}{\footnotesize $\tau = 0$} & \multicolumn{2}{c}{\footnotesize overlap} & \multicolumn{2}{c}{\footnotesize overlap} \\
    
    \cmidrule(rl){4-5} \cmidrule(rl){6-7} \cmidrule(rl){8-9} \cmidrule(rl){10-11} \cmidrule(rl){12-13}
    
     & {\bf\footnotesize Module} & {\bf\footnotesize Target} & {\bf\footnotesize OVRL} & {\bf\footnotesize XLSR} & {\bf\footnotesize OVRL} & {\bf\footnotesize XLSR}& {\bf\footnotesize OVRL} & {\bf\footnotesize XLSR} & {\bf\footnotesize OVRL} & {\bf\footnotesize XLSR} & {\bf\footnotesize OVRL} & {\bf\footnotesize XLSR} \\ 
    
    \toprule
    
    \multirow{6}{*}{\rotatebox{90}{CRUSE}} 
     & TAC  & rand & 2.02 & 2.08 & 2.09 & 2.24 & 2.13 & 2.27 & 2.05 & 2.23 & 2.01 & 2.06 \\
     & WCA  & rand & 2.19 & 2.08 & 2.25 & 2.25 & 2.31 & 2.31 & 2.24 & 2.27 & 2.19 & 2.09\\
    \cmidrule{2-13}
     & TAC  & min\_$\tau$ & 2.04 & 1.84 & 1.99 & 1.93 & 2.02 & 1.91 & 2.10 & 2.08 & 2.07 & 1.85 \\
     & WCA  & min\_$\tau$ & 2.23 & 2.32 & 2.18 & 2.26 & 2.21 & 2.32 & 2.35 & 2.76 & 2.27 & 2.39 \\
    \cmidrule{2-13}
     & TAC  & CM & 2.03 & 2.21 & 2.03 & 2.33 & 2.05 & 2.32 & 2.10 & 2.52 & 1.92 & 2.20 \\
     & WCA  & CM & \bf 2.43 & \bf 2.43 & \bf 2.51 & \bf 2.61 & \bf 2.52 & \bf 2.57 & \bf 2.53 & \bf 2.91 & \bf 2.44 & \bf 2.40 \\
    \midrule
    \multicolumn{3}{l}{\footnotesize Rand. Mic (Noisy)} & 1.70 & 2.11 & 1.61 & 2.18 & 1.63 & 2.14 & 1.72 & 2.29 & 1.61 & 2.13 \\
    \multicolumn{3}{l}{\footnotesize Single-ch + Rand. Mic} & 2.37 & 2.30 & 2.40 & 2.38 & 2.36 & 2.29 & 2.46 & 2.68 & 2.38 & 2.36 \\
    \multicolumn{3}{l}{\footnotesize Single-ch + Picknet} & 2.34 & 2.12 & 2.41 & 2.28 & 2.41 & 2.25 & 2.41 & 2.38 & 2.34 & 2.14 \\    
    \bottomrule
    \end{tabular}
    \end{center}
    \vspace{-1.2em}
\end{table*}

\subsection{Evaluation metrics}

In recent studies, MOS-based metrics have become the preferred evaluation method over intrusive measures like SDR, which often correlate less with actual perceived speech quality \cite{liu2024generative}.
Additionally, in our task, there is no single ideal target; as long as the model produces a clean mixture of active speakers, minor time delays do not significantly impact non-intrusive perceptual metrics, while they can lead to large discrepancies in intrusive metrics, which require strict alignment with a ground-truth reference.
To better capture perceptual speech quality, we evaluate our model using DNSMOS P.835 \cite{reddy2022dnsmos}, a DNN-based approach designed to approximate Mean Opinion Score (MOS) using the P.835 subjective evaluation framework. This includes three key metrics: speech quality (SIG), background noise quality (BAK), and overall quality (OVRL).
Additionally, we include XLS-R-MOS \cite{stahl2025distillation}, a large self-supervised model, fine-tuned on a diverse set of perceptual speech quality datasets, which showed superior generalization across various speech tasks, and can therefore be considered more reliable.
Lastly, we also include Cepstral Distance (CD) as an intrusive metric for additional analysis. 
However, CD values are target dependent, meaning direct comparisons across different targets may not be meaningful.

\subsection{Results}

Table \ref{tab:full-test-set} presents the evaluation results on the full test set using DNSMOS, XLS-R-MOS, and Cepstral Distance (CD) across different model configurations and target selection strategies. 
The comparison includes iFaSNet and CRUSE models, each incorporating either the TAC or WCA module, as well as baseline single-channel CRUSE models with random microphone selection and PickNet-based selection.
Across both iFaSNet and CRUSE, the WCA module consistently outperforms TAC in almost all configurations, demonstrating its effectiveness in asynchronous multi-device speech enhancement. 

Among all target selection strategies, using the closest microphone to each speaker (\textit{CM target}) achieves the best overall performance. 
With \textit{CM target}, applying WCA to CRUSE resulted in the highest OVRL (2.41) and XLSR (2.35) scores, alongside significant gains on iFaSNet.
This indicates that even though \textit{CM target} inherently requires speech separation within each microphone and microphone selection per speaker, the model with WCA learns well and performs strongly.
For random microphone target (\textit{rand target}), WCA shows consistent gains on CRUSE, while only a slight improvement over TAC when using iFaSNet.
This is likely due to the feature-level normalized cross-correlation module, which computes similarity scores relative to the target microphone (selected at random).
We retain this feature because it provides better performance compared to omitting it.
However, this approach inherently ties the model's output to a specific microphone order, making it sensitive to changes in microphone arrangement.
As a result, perturbing the microphone order could lead to significant variations in the model output, reducing its robustness in real-world multi-microphone scenarios.
This limitation can be mitigated by using the CRUSE model, which is order-agnostic and does not rely on a fixed microphone selection. 
Additionally, CRUSE outperforms iFaSNet in XLSR scores, further demonstrating its effectiveness.
Finally, under the minimum-latency target (\textit{min\_$\tau$ target}), WCA also consistently outperforms TAC across all metrics with CRUSE, especially on XLSR and CD score.
This shows that WCA is a versatile module that can be adapted to different targets based on user preference.

As baselines, we use (1) random mic selection (noisy), (2) a single-channel speech enhancement model (CRUSE) with random mic selection, and (3) single-channel CRUSE cascaded with Picknet \cite{yoshioka2022picknet}.
Although the single-channel CRUSE + PickNet achieves slight gains in BAK and OVRL scores, its XLSR score is notably lower compared to single-channel enhancement with random microphone selection.
We observe a discrepancy in microphone selection across frames when using PickNet, as its output is a weighted sum of different microphones at each frame. 
With the presence of device latency and clock drift, this inconsistency may reduce PickNet's effectiveness, potentially leading to degraded performance in real-world asynchronous multi-microphone scenarios.
In contrast, the WCA-equipped CRUSE model with the CM target effectively leverages and aggregates multi-microphone information, and outperforms these baselines. 
Overall, these results demonstrate that WCA improves speech quality more effectively than TAC, particularly in CRUSE, and our training target selection strategies further improve performance, with the closest-microphone target consistently achieving the best results.

We also conduct an ablation study on five different conditions, where we create new test sets varying clock drift standard deviation ($\gamma)$, introducing always a maximum delay of $\Delta\tau = 40$ms instead of Gaussian distributed delays, and creating non-overlapping and fully-overlapping multi-talk. 
The results are shown for CRUSE only in Table \ref{tab:ablation}.
Among these conditions, the most challenging scenarios occur when speech is fully overlapped or when there is a maximum device latency difference of 40 ms, both of which significantly impact performance.
In contrast, the model performs exceptionally well on non-overlapping speech, achieving the highest OVRL score of 2.53 and XLSR score of 2.91 with CRUSE + WCA using the CM target.
Additionally, the model demonstrates robustness to clock drift, as both a slight drift (std($\gamma$) = 0.5) and a stronger drift (std($\gamma$) = 2.0) have minimal impact on performance, indicating its stability under varying synchronization conditions.
Audio examples applied to real recordings are available at \url{https://distributedse.github.io/waspaa2025-93/}.
\section{Conclusion}

We propose Windowed Cross-Attention (WCA), an efficient temporal cross-attention module designed to effectively aggregate multi-microphone signals under challenging conditions, where device latency and clock drift are both present and unknown. 
WCA is designed to handle real-time continuous streaming multi-microphone speech input, offering adaptive alignment capabilities while supporting arbitrary microphone geometries and varying numbers of speakers.
Our results demonstrate that WCA consistently outperforms previous methods and strong baseline models. 
Notably, WCA demonstrates strong generalization across different conditions, particularly when trained with the mixture of the closest microphones to each active speaker.
The improvements observed across multiple configurations confirm that WCA is a powerful and flexible solution for asynchronous multi-microphone speech enhancement in real-world scenarios.

\clearpage
\IEEEtriggeratref{22}

\bibliographystyle{IEEEtran}
\bibliography{mybib}

\begin{thebibliography}{10}
\providecommand{\url}[1]{#1}
\csname url@samestyle\endcsname
\providecommand{\newblock}{\relax}
\providecommand{\bibinfo}[2]{#2}
\providecommand{\BIBentrySTDinterwordspacing}{\spaceskip=0pt\relax}
\providecommand{\BIBentryALTinterwordstretchfactor}{4}
\providecommand{\BIBentryALTinterwordspacing}{\spaceskip=\fontdimen2\font plus
\BIBentryALTinterwordstretchfactor\fontdimen3\font minus
  \fontdimen4\font\relax}
\providecommand{\BIBforeignlanguage}[2]{{%
\expandafter\ifx\csname l@#1\endcsname\relax
\typeout{** WARNING: IEEEtran.bst: No hyphenation pattern has been}%
\typeout{** loaded for the language `#1'. Using the pattern for}%
\typeout{** the default language instead.}%
\else
\language=\csname l@#1\endcsname
\fi
#2}}
\providecommand{\BIBdecl}{\relax}
\BIBdecl

\bibitem{gaubitch2013auto}
N.~D. Gaubitch, W.~B. Kleijn, and R.~Heusdens, ``Auto-localization in ad-hoc
  microphone arrays,'' in \emph{ICASSP}, 2013.

\bibitem{wang2015self}
L.~Wang, T.-K. Hon, J.~D. Reiss, and A.~Cavallaro, ``Self-localization of
  ad-hoc arrays using time difference of arrivals,'' \emph{IEEE Transactions on
  Signal Processing}, vol.~64, 2015.

\bibitem{cherkassky2017blind}
D.~Cherkassky and S.~Gannot, ``Blind synchronization in wireless acoustic
  sensor networks,'' \emph{IEEE/ACM Transactions on Audio, Speech, and Language
  Processing}, 2017.

\bibitem{bahari2017blind}
M.~H. Bahari, A.~Bertrand, and M.~Moonen, ``Blind sampling rate offset
  estimation for wireless acoustic sensor networks through weighted
  least-squares coherence drift estimation,'' \emph{IEEE/ACM Transactions on
  Audio, Speech, and Language Processing}, 2017.

\bibitem{hu2022distributed}
D.~Hu, H.~Zhang, F.~Bao, and R.~Wang, ``Distributed sampling rate offset
  estimation over acoustic sensor networks based on asynchronous network newton
  optimization,'' \emph{IEEE/ACM Transactions on Audio, Speech, and Language
  Processing}, 2022.

\bibitem{schmalenstroeer2017multi}
J.~Schmalenstroeer, J.~Heymann, L.~Drude, C.~Boeddecker, and R.~Haeb-Umbach,
  ``Multi-stage coherence drift based sampling rate synchronization for
  acoustic beamforming,'' in \emph{2017 IEEE 19th International Workshop on
  Multimedia Signal Processing (MMSP)}.\hskip 1em plus 0.5em minus 0.4em\relax
  IEEE, 2017.

\bibitem{9003849}
Y.~Luo, C.~Han, N.~Mesgarani, E.~Ceolini, and S.-C. Liu, ``Fasnet: Low-latency
  adaptive beamforming for multi-microphone audio processing,'' in \emph{2019
  IEEE Automatic Speech Recognition and Understanding Workshop (ASRU)}, 2019.

\bibitem{9054177}
Y.~Luo, Z.~Chen, N.~Mesgarani, and T.~Yoshioka, ``End-to-end microphone
  permutation and number invariant multi-channel speech separation,'' in
  \emph{ICASSP}, 2020.

\bibitem{luo21c_interspeech}
Y.~Luo and N.~Mesgarani, ``Implicit filter-and-sum network for end-to-end
  multi-channel speech separation,'' in \emph{Interspeech}, 2021.

\bibitem{wang2018multichannel}
Z.-Q. Wang, J.~Le~Roux, and J.~R. Hershey, ``Multi-channel deep clustering:
  Discriminative spectral and spatial embeddings for speaker-independent speech
  separation,'' in \emph{ICASSP}, 2018.

\bibitem{kim24m_interspeech}
J.~Kim, S.~Kindt, N.~Madhu, and H.-G. Kang, ``Enhanced deep speech separation
  in clustered ad hoc distributed microphone environments,'' in
  \emph{Interspeech}, 2024.

\bibitem{zhang2021}
X.-L. Zhang, ``Deep ad-hoc beamforming,'' \emph{Computer Speech \& Language},
  vol.~68, 2021.

\bibitem{9746876}
T.~Yoshioka, X.~Wang, D.~Wang, M.~Tang, Z.~Zhu, Z.~Chen, and N.~Kanda,
  ``Vararray: Array-geometry-agnostic continuous speech separation,'' in
  \emph{ICASSP}, 2022.

\bibitem{pandey22c_interspeech}
A.~Pandey, B.~Xu, A.~Kumar, J.~Donley, P.~Calamia, and D.~Wang, ``Time-domain
  ad-hoc array speech enhancement using a triple-path network,'' in
  \emph{Interspeech}, 2022.

\bibitem{6694323}
R.~Sakanashi, N.~Ono, S.~Miyabe, T.~Yamada, and S.~Makino, ``Speech enhancement
  with ad-hoc microphone array using single source activity,'' in \emph{2013
  Asia-Pacific Signal and Information Processing Association Annual Summit and
  Conference}, 2013.

\bibitem{tavakoli2016framework}
V.~M. Tavakoli, J.~R. Jensen, M.~G. Christensen, and J.~Benesty, ``A framework
  for speech enhancement with ad hoc microphone arrays,'' \emph{IEEE/ACM
  Transactions on Audio, Speech, and Language Processing}, vol.~24, 2016.

\bibitem{capon1969high}
J.~Capon, ``High-resolution frequency-wavenumber spectrum analysis,''
  \emph{Proceedings of the IEEE}, vol.~57, 1969.

\bibitem{griffiths1982alternative}
L.~Griffiths and C.~Jim, ``An alternative approach to linearly constrained
  adaptive beamforming,'' \emph{IEEE Transactions on antennas and propagation},
  vol.~30, 1982.

\bibitem{souden2009optimal}
M.~Souden, J.~Benesty, and S.~Affes, ``On optimal frequency-domain multichannel
  linear filtering for noise reduction,'' \emph{IEEE Transactions on audio,
  speech, and language processing}, vol.~18, 2009.

\bibitem{gannot2017consolidated}
S.~Gannot, E.~Vincent, S.~Markovich-Golan, and A.~Ozerov, ``A consolidated
  perspective on multimicrophone speech enhancement and source separation,''
  \emph{IEEE/ACM Transactions on Audio, Speech, and Language Processing},
  vol.~25, 2017.

\bibitem{araki2018}
S.~Araki, N.~Ono, K.~Kinoshita, and M.~Delcroix, ``Comparison of reference
  microphone selection algorithms for distributed microphone array based speech
  enhancement in meeting recognition scenarios,'' in \emph{2018 16th
  International Workshop on Acoustic Signal Enhancement (IWAENC)}, 2018.

\bibitem{kumatani2011channel}
K.~Kumatani, J.~McDonough, J.~F. Lehman, and B.~Raj, ``Channel selection based
  on multichannel cross-correlation coefficients for distant speech
  recognition,'' in \emph{2011 Joint Workshop on Hands-free Speech
  Communication and Microphone Arrays}.\hskip 1em plus 0.5em minus 0.4em\relax
  IEEE, 2011.

\bibitem{wolf2014channel}
M.~Wolf and C.~Nadeu, ``Channel selection measures for multi-microphone speech
  recognition,'' \emph{Speech Communication}, vol.~57, 2014.

\bibitem{zhang2020study}
J.~Zhang, H.~Chen, L.-R. Dai, and R.~C. Hendriks, ``A study on reference
  microphone selection for multi-microphone speech enhancement,''
  \emph{IEEE/ACM Transactions on Audio, Speech, and Language Processing},
  vol.~29, 2020.

\bibitem{yoshioka2022picknet}
T.~Yoshioka, X.~Wang, and D.~Wang, ``Picknet: Real-time channel selection for
  ad hoc microphone arrays,'' in \emph{ICASSP}, 2022.

\bibitem{wang2021continuous}
D.~Wang, T.~Yoshioka, Z.~Chen, X.~Wang, T.~Zhou, and Z.~Meng, ``Continuous
  speech separation with ad hoc microphone arrays,'' in \emph{2021 29th
  European Signal Processing Conference (EUSIPCO)}.\hskip 1em plus 0.5em minus
  0.4em\relax IEEE, 2021.

\bibitem{8234698}
J.~Zhang, S.~P. Chepuri, R.~C. Hendriks, and R.~Heusdens, ``Microphone subset
  selection for mvdr beamformer based noise reduction,'' \emph{IEEE/ACM
  Transactions on Audio, Speech, and Language Processing}, 2018.

\bibitem{zhang2021sensor}
J.~Zhang, J.~Du, and L.-R. Dai, ``Sensor selection for relative acoustic
  transfer function steered linearly-constrained beamformers,'' \emph{IEEE/ACM
  Transactions on Audio, Speech, and Language Processing}, 2021.

\bibitem{miyabe2015blind}
S.~Miyabe, N.~Ono, and S.~Makino, ``Blind compensation of interchannel sampling
  frequency mismatch for ad hoc microphone array based on maximum likelihood
  estimation,'' \emph{Signal Processing}, vol. 107, 2015.

\bibitem{yoshioka2019meeting}
T.~Yoshioka, D.~Dimitriadis, A.~Stolcke, W.~Hinthorn, Z.~Chen, M.~Zeng, and
  X.~Huang, ``Meeting transcription using asynchronous distant microphones.''
  in \emph{Interspeech}, 2019.

\bibitem{furnon2023samba}
N.~Furnon, R.~Serizel, S.~Essid, and I.~Illina, ``Samba: Speech enhancement
  with asynchronous ad-hoc microphone arrays,'' \emph{arXiv preprint
  arXiv:2307.16582}, 2023.

\bibitem{guo24_interspeech}
H.~Guo, Y.~Chen, X.-L. Zhang, and X.~Li, ``Graph attention based multi-channel
  u-net for speech dereverberation with ad-hoc microphone arrays,'' in
  \emph{Interspeech}, 2024.

\bibitem{ristea23_interspeech}
N.~C. Ristea, E.~Indenbom, A.~Saabas, T.~Pärnamaa, J.~Guzhvin, and R.~Cutler,
  ``Deepvqe: Real time deep voice quality enhancement for joint acoustic echo
  cancellation, noise suppression and dereverberation,'' in \emph{Interspeech},
  2023.

\bibitem{9413580}
S.~Braun, H.~Gamper, C.~K. Reddy, and I.~Tashev, ``Towards efficient models for
  real-time deep noise suppression,'' in \emph{ICASSP}, 2021.

\bibitem{10096131}
J.~Neri and S.~Braun, ``Towards real-time single-channel speech separation in
  noisy and reverberant environments,'' in \emph{ICASSP}, 2023.

\bibitem{dnschallenge2023}
H.~Dubey, A.~Aazami, V.~Gopal, B.~Naderi, S.~Braun, R.~Cutler, A.~Ju,
  M.~Zohourian, M.~Tang, M.~Golestaneh, and R.~Aichner, ``Icassp 2023 deep
  noise suppression challenge,'' \emph{IEEE Open Journal of Signal Processing},
  vol.~5, 2024.

\bibitem{Allen1979}
J.~B. Allen and D.~A. Berkley, ``Image method for efficiently simulating
  small-room acoustics,'' \emph{The Journal of the Acoustical Society of
  America}, vol.~65, 1979.

\bibitem{luo2020dual}
Y.~Luo, Z.~Chen, and T.~Yoshioka, ``Dual-path rnn: efficient long sequence
  modeling for time-domain single-channel speech separation,'' in
  \emph{ICASSP}, 2020.

\bibitem{Braun2022}
S.~Braun and H.~Gamper, ``Effect of noise suppression losses on speech
  distortion and {ASR} performance,'' in \emph{ICASSP}, 2022.

\bibitem{braun2021consolidated}
S.~Braun and I.~Tashev, ``A consolidated view of loss functions for supervised
  deep learning-based speech enhancement,'' in \emph{2021 44th International
  Conference on Telecommunications and Signal Processing (TSP)}, 2021.

\bibitem{liu2024generative}
A.~H. Liu, M.~Le, A.~Vyas, B.~Shi, A.~Tjandra, and W.-N. Hsu, ``Generative
  pre-training for speech with flow matching,'' in \emph{ICLR}, 2024.

\bibitem{reddy2022dnsmos}
C.~K. Reddy, V.~Gopal, and R.~Cutler, ``Dnsmos p. 835: A non-intrusive
  perceptual objective speech quality metric to evaluate noise suppressors,''
  in \emph{ICASSP}, 2022.

\bibitem{stahl2025distillation}
B.~Stahl and H.~Gamper, ``Distillation and pruning for scalable self-supervised
  representation-based speech quality assessment,'' \emph{ICASSP}, 2025.

\end{thebibliography}

\end{document}